\documentclass[11pt]{article}
\usepackage{amssymb}
\usepackage{graphicx}
\def\@oddhead{}\def\@evenhead{}
\def\@oddfoot{}
\def\@evenfoot{\@oddfoot}
\begin{document}
\begin{flushright}
{\bf ATLAS Internal Note \\
INDET-NO-023\\
February 9, 1993}
\end{flushright}
\vskip 0.7cm
\begin{center}
{\Large\bf {Field integrals for the ATLAS tracking volume}}\\[7mm]
{\bf\normalsize {V.I. Klyukhin}\\[2mm]
{\em {IHEP, Protvino, Russia}}\\[4mm]
{A. Poppleton}\\[2mm]
{\em {CERN, Geneva, Switzerland}}\\[4mm]
{J. Schmitz}\\[2mm]
{\em {NIKHEF-H, Amsterdam, The Netherlands}}}\\[4mm]
\end{center}
\vspace{0.7cm}
\section {Introduction}

The ATLAS inner tracker measures charged track momenta from the deflection in a 
solenoidal magnetic field. The choice of a long tracking cavity, allied to the 
decision to build the coil inside the electromagnetic calorimeter, leads to a 
favourable solenoid aspect ratio which ensure a highly uniform field in the 
central tracking region. However the situation is not ideal for tracks with 
$|\eta| \gtrsim 1.8$, since $\sim$25~cm before the end of the coil the magnetic 
field lines become parallel to the direction of stiff tracks from the vertex 
region. The subsequent track deflection is in the opposite sence to that 
experienced in the central region. As the last tracking detector lies about 
20~cm beyond the end of the coil, almost 15\% of the length of these forward 
tracks is subject to reverse bending. The purpose of this note is to compare 
the expected momentum resolution from the LOI magnet configuration with that 
from an `ideal' (infinity long) solenoid with constant field in the axial 
direction. 

The calculations were performed using a field map extracted from the POISSON 
program package \cite{Holsinger}. The solenoidal coil was set to be 6.3~m long 
with a radius of 1.23~m. The number of Ampere-turns of the coil was tuned to 
give a magnetic flux density of 2~T at the centre of the solenoid. The 
cylindrical inner cavity for tracking detectors was taken to be 6.8~m long with 
1.06~m radius. The field lines (assuming an iron return flux structure outside 
the hadronic calorimeter) are illustrated in Fig.~1. Fig.~2 and Fig.~3 show the 
variation of the magnetic flux density across radial and logitudinal sections 
of the tracking volume, respectively.
\section {Magnetic field integrals and momentum resolution}

To derive the connection between the momentum resolution of a particle 
passing through this solenoid and the field integral over the track length, 
concider the trajectory of a particle emitted, at angle $\theta$ to the beam 
axis, from the nominal beam crossing point (which is taken as the origin of the 
coordinate system). The $z$-direction is defined to be along the beam axis, and 
the transverse radius $r$ is the orthogonal distance from the $z$-axis.

For a small step $d\vec l$ along the direction of the particle motion in an 
ideal solenoid, $d\alpha$ (the change to the turning angle of the track) lies 
in the transverse plane and is given by

\begin{equation}
d\alpha = \frac{0.3}{p_T} B dl \sin{\theta},
\end{equation}

\noindent where $l$ is in meters, $p_T$ is the transverse momentum of the 
particle in GeV/$c$, and $\vec B$ is the vector of the magnetic flux density in 
Tesla. In general, for an inhomogeneous field, the track is turned according 
to, and in the direction of, the vector product $d\vec l \times \vec B$. 
However for tracks from the vertex region of the ATLAS solenoid, the majority 
of the tracking volume is contained in a good approximation to the field from 
an ideal solenoid, and furthermore $\vec B$ and $d\vec l$ are nearly parallel 
for the remaining trajectory (which is at large $|z|$). Thus the subsequent 
comparison is restricted to the transverse projection; firstly because it 
contains the dominant deflection component, and secondly because precision 
measurements to determine the track momentum will probably only be available 
in this plane.

For energetic particles the magnetic deflection is small compared to the track 
length, thus the distance along the trajectory can be approximated by 
$l = r/\sin{\theta}$ and small angle approximations are valid. At $l$ the 
relative angle $\alpha$ of the track with respect to its initial direction in 
transverse projection is given by

\begin{equation}
\alpha\,(l) = \frac{0.3}{p_T} \int \limits_{0}^{l}B \sin{\theta_{(d\vec l, \vec B)}} dl,
\end{equation}

\noindent where $\theta_{(d\vec l, \vec B)}$ represents the longitudinal 
component of the angle between the track and field vectors. The toal transverse 
deflection $x$ is obtained by integrating eq.(2) over $dr = dl \sin{\theta}$:

\begin{equation}
x\,(l) = \frac{0.3}{p_T} \int \limits_{0}^{l\sin{\theta}}\, 
\int \limits_{0}^{r/\sin{\theta}}B 
\sin{\theta_{(d\vec l, \vec B)}} dl dr.
\end{equation}

\noindent For the ideal solenoid, $x(l)$ is simply proportional to $l^2$.

Now consider a cluster of measurements, near the end of the tracking volume, 
providing a momentum measurement by their impact parameter at the beam axis. In 
this limiting case, the relative momentum precision for a real versus ideal 
solenoid is given by $R$, the relative values of $x(L)$, where $L$ is the total 
track length in the tracking volume. The degradation in $\Delta p/p$ is thus 
$1 - R$.

However a more precise momentum measurement can be obtained from measuring the 
track sagitta, which is $x(L)/2 - x(L/2)$. For the ideal solenoid the sagitta 
equals $x(L/2)$ since $x(L) = 4x(L/2)$ from eq.(3). The value of $x(L/2)$ is 
essentially the same for the ideal and ATLAS solenoids, for the reason given 
previously, so the ratio of the sagittas is given by $2R - 1$ and $\Delta p/p$ 
is degraded by $2(1 - R)$. The sagitta measurement is the least favourable 
extreme when taken in comparison with the ideal solenoid; to give an example: 
if the relative double field integral $R$ equals 0.85, then $\Delta p/p$ is 
degraded by 15\% for an impact parameter measurement and 30\% if the sagitta is 
measured.

In practice, measurements are taken along the track length causing degradation 
which lies between these two extremes; however the current ATLAS forward 
detector layout is close to the sagitta extreme, so $\Delta p/p$ is degraded by 
almost twice as much as the field integral.

\newpage
\vskip 1.4cm
\begin{center}
\includegraphics[width=0.95\textwidth]{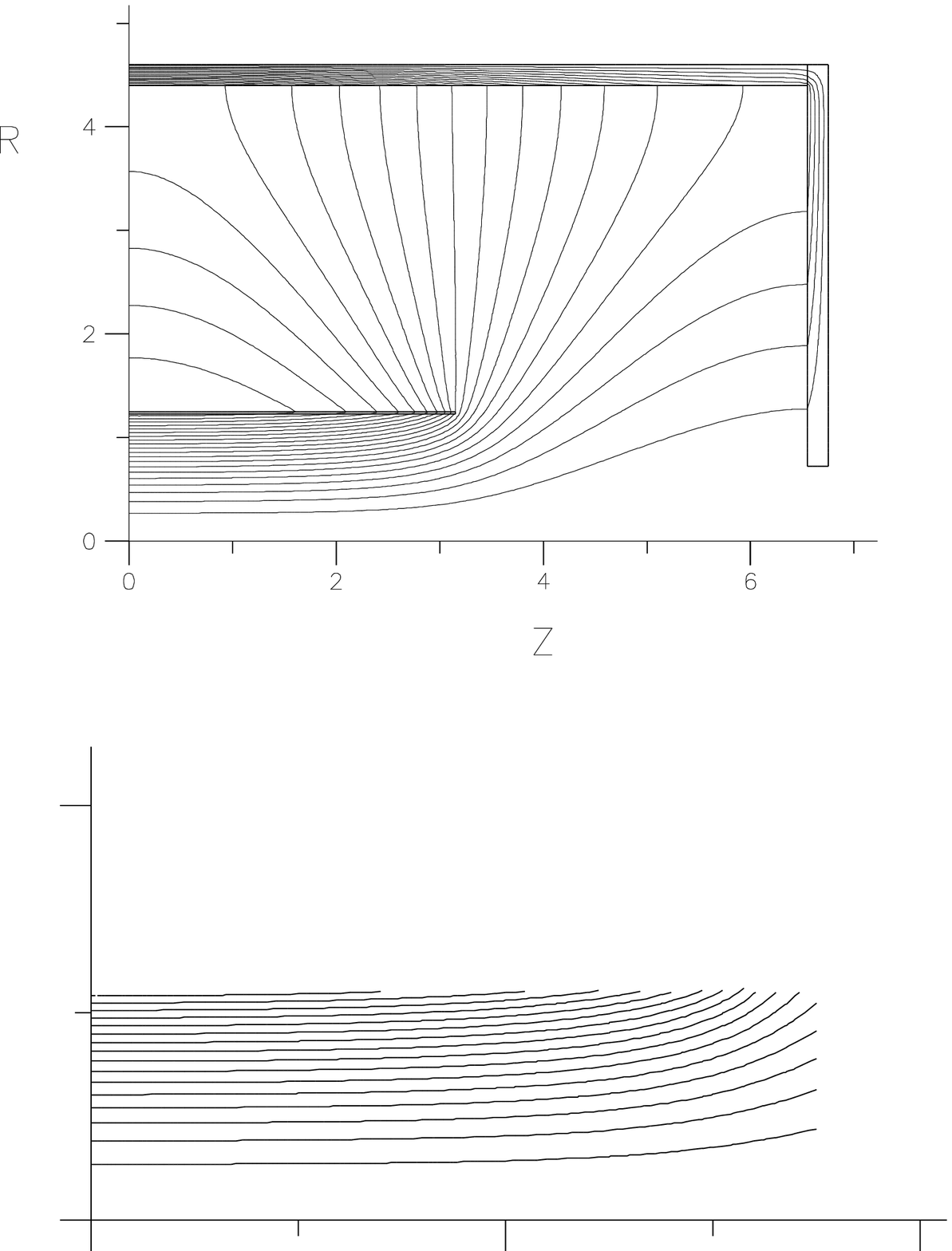}
\end{center}
\vskip 0.7cm
\noindent Figure 1: top) {\em Field lines for ATLAS solenoid with iron return 
flux}, \hspace{1cm} bottom) {\em zoom onto inner tracking cavity}.

\newpage
\vskip 1.4cm
\begin{center}
\includegraphics[width=0.6\textwidth]{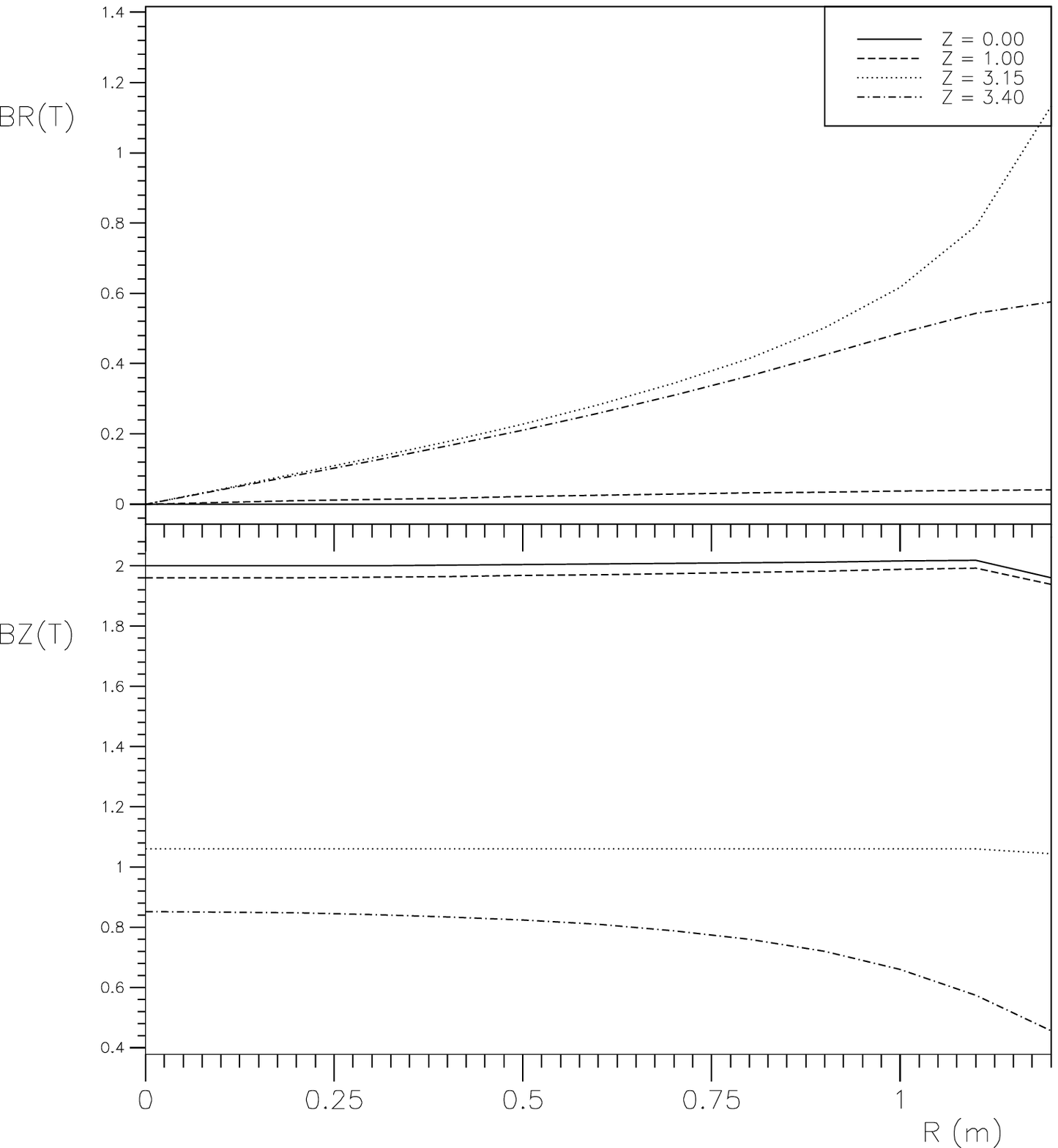}

\noindent Figure 2: {\em Field components across radial sections of the 
tracking volume}.
\end{center}
\begin{center}
\includegraphics[width=0.6\textwidth]{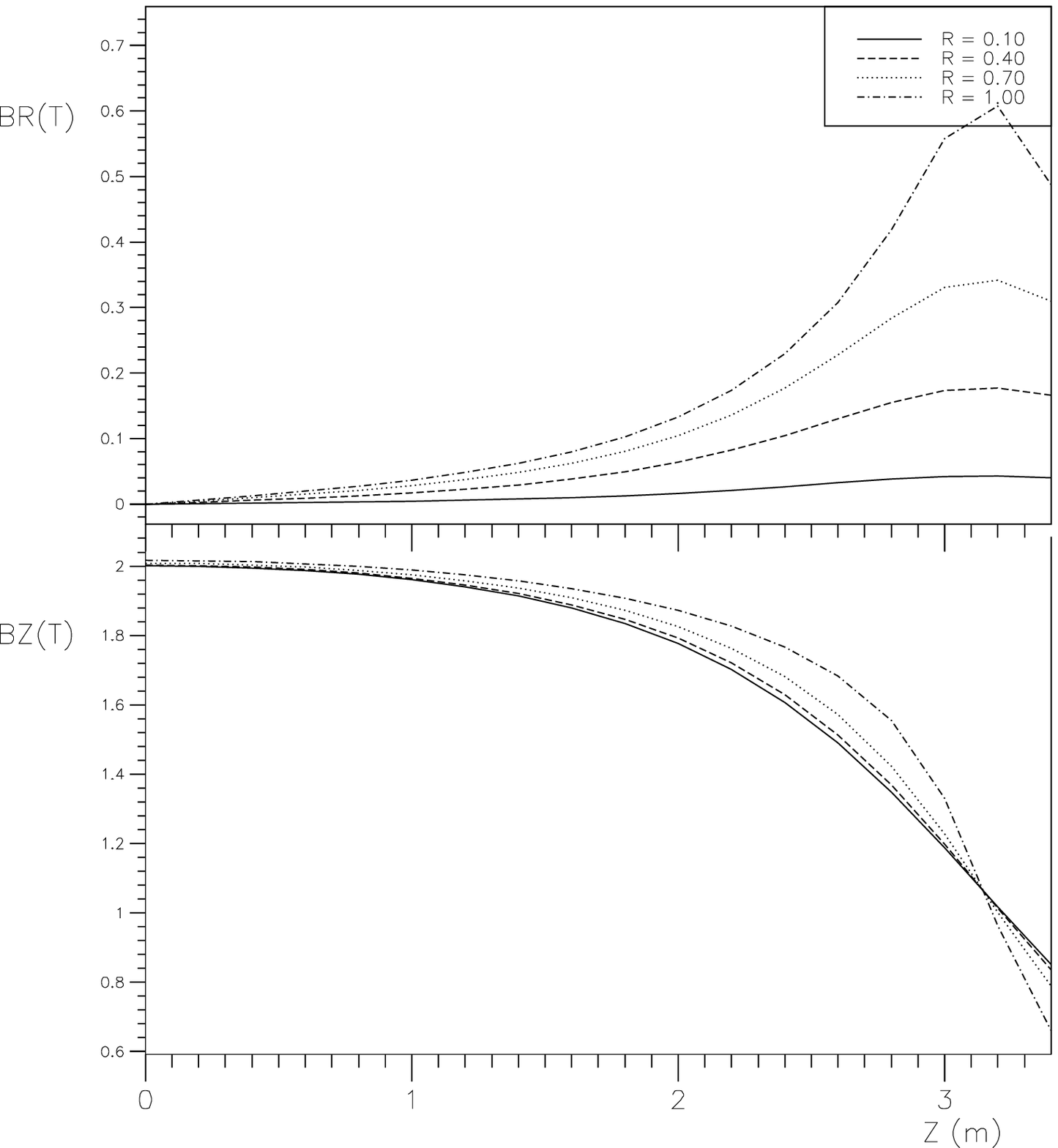}

\noindent Figure 3: {\em Field components across longitudinal section of the 
tracking volume}.
\end{center}
\newpage
\vskip 1.4cm
\begin{center}
\includegraphics[width=0.6\textwidth]{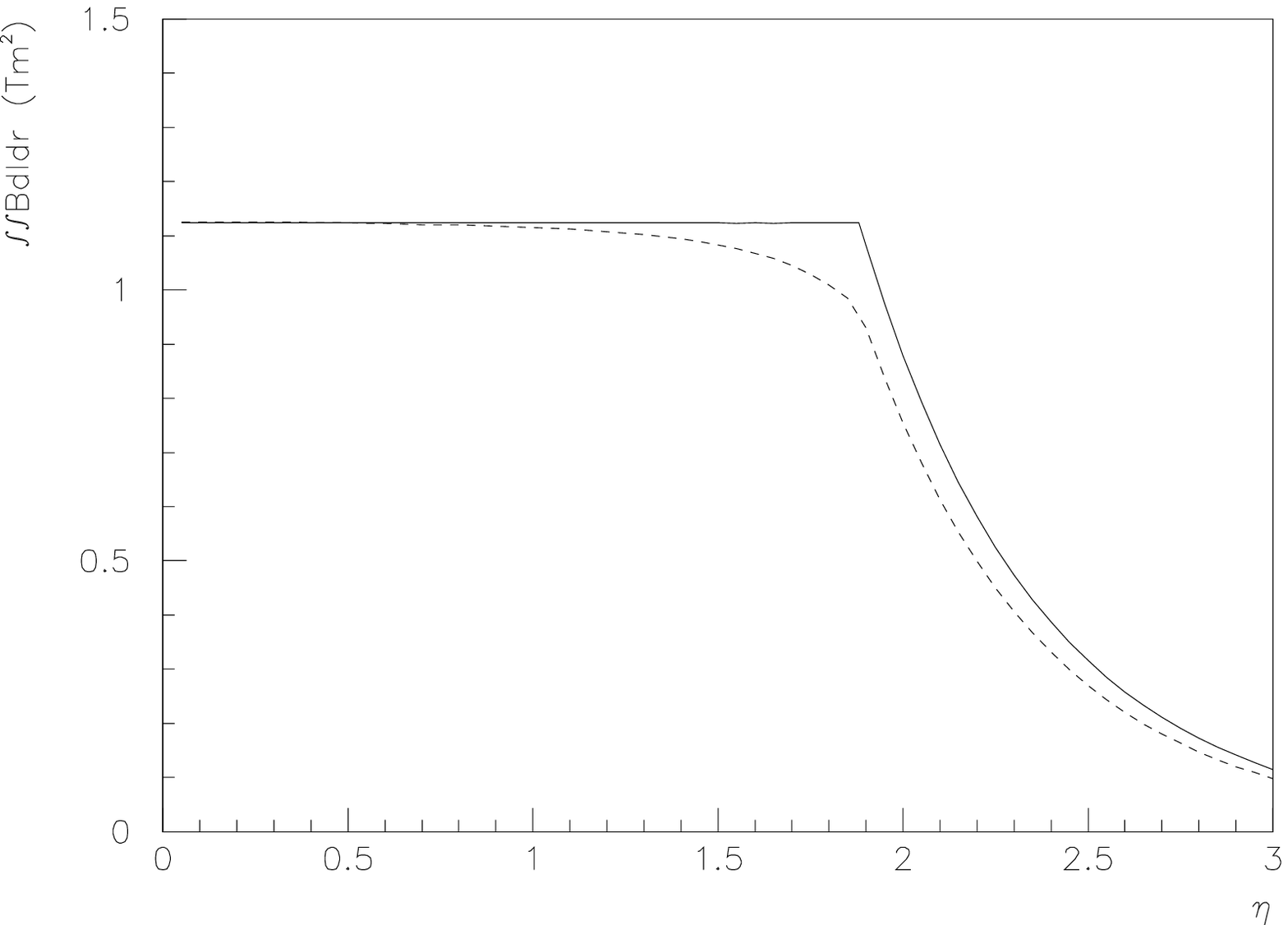}
\end{center}

\noindent Figure 4: {\em The double field integral $I_2$ for the ideal (solid 
line) and inhomogeneous (dashed line) cases}. 

\section {Results}

To evaluate realistic ATLAS (inhomogeneous) fields integrals, the program 
POISGT \cite{Klyukhin} was used to extract magnetic field map from POISSON. 
This field map is meant to be linked to DICE for accurate simulations. The 
dependence of the double integral

\begin{equation}
I_2 = \int \limits_{0}^{L\sin{\theta}}\,\int \limits_{0}^{r/\sin{\theta}}B 
\sin{\theta_{(d\vec l, \vec B)}} dl dr
\end{equation}

\noindent versus $\eta$ is shown in Fig.4 for the ideal (solid line) and 
inhomogeneous (dashed line) fields. Fig.5 shows $(1 - R)$, the degradation in 
this field integral, as a function of $\eta$.

It can be seen that $I_{2_i}$ ($I_2$ inhomogeneous) is almost constant up to 
$|\eta| =$ 0.6 (it decreases by only 0.1\% when compared to the ideal 
solenoid); the decrease reaches 1\% by $|\eta| =$ 1.1, 5\% at $|\eta| =$ 1.6 
and 10\% at $|\eta| =$ 1.8. Beyond the corner of the inner tracking cavity in 
the $rz$-plane ($|\eta| >$ 1.88), the degradation only increases slowly from 
$\sim$14\% at $|\eta| =$ 1.88 to $\sim$15\% at $|\eta| =$ 3.0. Of course the 
magnitude of the field integral is falling rapidly in this region due to the 
reduced radial distance traversed. This directly effects the momentum 
resolution at a given $p_T$ according to eq.(3). \footnote{However there is an 
additional $\sin{\theta}$ term to be included when considering the momentum 
resolution for a fixed value of $p$, which is perhaps more appropriate for 
extremely high momenta. From this viewpoint $x(L)$, the field integral 
contribution to $\Delta p/p$, rises from $\eta =$ 0 to a maximum at $|\eta| =$ 
1.8, then falls back to approximately the $\eta =$ 0 value at $|\eta| =$ 3, 
i.e. although $I_{2_i}$($\eta =$ 3) $\sim 0.1 I_{2_i}$($\eta =$ 0), a 1 TeV 
track at $|\eta| =$ 3 has $p_T \sim 100$~GeV, thus much the same $\Delta p/p$ 
as for 1 TeV at $\eta =$ 0 under the assumption of similarly precise measuring 
stations.}  

An important question is whether the extra lever arm at the end of the tracking 
cavity is significantly improving the momentum resolution in this region 
($|\eta| >$ 1.88), since some free space in front of the end-caps might be 
desirable for installation convenience (e.g. to provide a patch-pannel) and 
will be needed for the detector frames. For the ideal solenoid, moving the last 
chamber inwards by 20 cm would reduce the field integral by 11.4\%; with the 
inhomogeneous field this reduction is less pronounced ($\sim$7\%) but still 
significant, the $1 - R$ degradation decreases accordingly.

\vskip 1.4cm
\begin{center}
\includegraphics[width=0.6\textwidth]{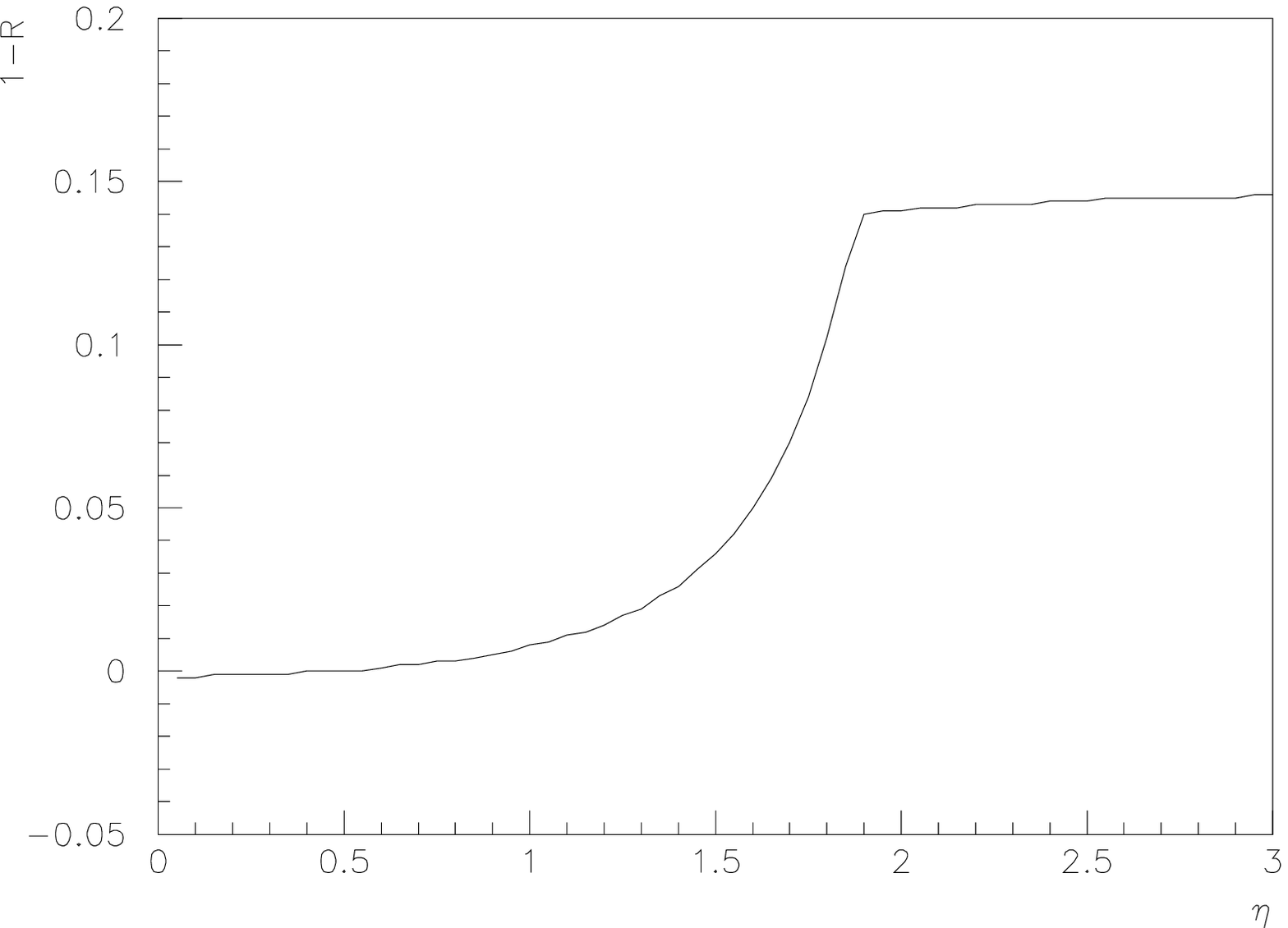}

\noindent Figure 5: {\em The field integral degradation as a function of 
$\eta$}.
\end{center}

To conclude, the current geometry of the inner cavity of the ATLAS detector 
results in an ideal 2 T magnetic field for ($|\eta| <$ 0.6. Beyond that 
rapidity, there is a loss of bending power which rises to a plateau for 
$|\eta| \gtrsim$ 1.9. Depending on the $z$-coordinate of the last tracking 
chamber, the field integral degradation for $|\eta| \gtrsim$ 1.9 is 
$\sim 10 - 15$\%, resulting in a $20 - 30$\% worsening of the $p_T$-resolution 
with respect to the value that would be obtained from an ideal solenoid field.


\begin{thebibliography}{99}
\bibitem{Holsinger}
R.F. Holsinger and C. Iselin. 
{\em The CERN-POISSON Program Package (POISCR) User Guide.} 
CERN Computer Center Program Library, T604, (1984).
\bibitem{Klyukhin}
V.I. Klyukhin.
{\em Interface Program for POISSON and GEANT Packages.}
Proc. of the Conf. on Computing in High Energy Physics, 21-25 September 1992,
Annecy, France.
\end{thebibliography}
\end{document}